\documentclass{sf2a-conf2012}
\usepackage{graphicx}
\usepackage[]{natbib}  
\usepackage[cyr]{aeguill}

\def\BibTeX{{\rm B\kern-.05em{\sc i\kern-.025em b}\kern-.08em
    T\kern-.1667em\lower.7ex\hbox{E}\kern-.125emX}}
\bibpunct{(}{)}{;}{a}{}{,}  
\newcommand{\numax}{{\nu_\mathrm{max}}}
\newcommand{\nuc}{{\nu_\mathrm{c}}}

\begin{document}

\TitreGlobal{SF2A 2012}


\title{Determination of the stars fundamental parameters using seismic scaling relations}

\runningtitle{Determination of the star fundamental parameters using seismic scaling relations}

\author{K. Belkacem}
\address{LESIA, UMR8109, Observatoire de Paris, Universit\'e Pierre et Marie Curie, 
Universit\'e Denis Diderot, 92195 Meudon Cedex, France. E-mail: \emph{kevin.belkacem@obspm.fr} }

\setcounter{page}{237}


\maketitle


\begin{abstract}
Seismology of stars that exhibit solar-like oscillations develops a growing interest with the wealth of observational results obtained with the CoRoT and {\it Kepler} space-borne missions. In this framework,  relations between asteroseismic quantities and stellar parameters provide a unique opportunity to derive model-independent determinations of stellar parameters (e.g., masses and radii) for a large sample of stars. I  review those scaling relations with particular emphasis on the underlying physical processes governing those relations, as well as their uncertainties. 
 \end{abstract}

\begin{keywords}
Convection - Turbulence - Stars: oscillations - Stars: interiors
\end{keywords}

\section{Introduction}

The determination of accurate stellar parameters (mass, radius, effective temperature, age, and chemical composition) is a fundamental and longstanding problem in astrophysics \citep[e.g.][]{Soderblom2010}. When compared to many other methods, seismology is recognized to provide the most precise determination of those fundamental parameters \citep[e.g.][]{Lebreton2011}. 
Nevertheless, such a determination is only possible by means of the use of stellar models and therefore suffers from our deficient knowledge of the physical processes taking place in stars \citep[e.g.][]{Goupil2011a,Goupil2011b}. 

The situation recently improved with the advent of space-borne asteroseismology, more precisely with the launch of CoRoT \citep{Baglin2006a,Baglin2006b,Michel2008} and {\it Kepler} \citep{Borucki2010}. Those two spacecrafts are providing us with high-quality photometric data. Up to now, several hundreds of main-sequence stars with solar-like oscillations have been detected and several thousands oscillating red-giant stars, allowing for statistical analysis. With such large number of stars it is not possible to performe classical seismology, \emph{i.e.} by individual mode fitting of the power spectrum. This is very time (and man-power) consuming so that a new method emerged through the use of seismic global parameters. The latter are typical global characteristics of the oscillation spectra such that the regularities in frequency (or period), or the frequency of the maximum amplitude. 

This approach gave birth to the \emph{ensemble} asteroseismology, which cornerstones are the relations between global seismic quantities and stellar parameters. It allows ones to infer model-independent stellar parameters as well as information on stellar structure and evolution. 
Scaling relations  between asteroseismic quantities  and stellar  parameters such as stellar mass, radius, effective temperature, and luminosity  have initially been observationally  derived  by several authors \citep[e.g.][]{Ulrich86,Brown91,Kjeldsen95} using ground-based data. 
CoRoT and Kepler confirmed these results by providing accurate and homogeneous measurements for a large sample of stars from main-sequence to red-giant stars \citep[e.g., ][]{Mosser2010,Baudin2011a,Mosser2011,Mosser2011b,Mosser2012b,Mosser12a,Samadi2012}.  

Among them, the relations between the large separation ($\Delta \nu$) and the mean density ($\bar{\rho}$) as well as between the frequency of the maximum height in the power spectrum ($\numax$) and the photospheric cut-off frequency ($\nuc$) have been known for a long time in the context of solar-like pulsators \citep[e.g.,][]{Ulrich86,Brown91,KB95,Belkacem2011}. In fact, as shown below, the physical grounds of these two relations were known for an even longer time in the context of classical pulsators (stars exhibiting opacity-driven modes). This will be the subject of the first part of this review, with particular emphasize on the theoretical side. 
We will discuss the fundamental physical concept underlying these scaling relations and show how one can use them to infer stellar masses and radii. Section 3 will be dedicated to scaling relations that exhibit an important potential in providing information on the core properties of stars, or effective temperature. Last but not least, we will discuss the uncertainties on the seismic global parameters by showing that many biases can exist and are still to be addressed. 

Finally, we note that this paper is by no means an exhaustive review. Rather it puts forth the necessity of understanding the theoretical ground of the scaling relations, unfortunately too often bypassed in the recent literature. 

\section{The canonical scaling relations}

In this section, we will first address the problem of the relations between the large separation ($\Delta \nu$) and the mean density ($\bar{\rho}$), as well as the frequency of maximum height in the power spectrum ($\nu_{\rm max}$) and the cut-off frequency ($\nu_c$). Those two scaling relations provide an estimate of the mass and the radius. They are now widely used and we will see that the underlying physics is known for a long time. Consequently, in this review we will denote them as \emph{canonical} scaling relations. 

\subsection{Relation between the large separation ($\Delta \nu$) and the mean density ($\bar{\rho}$)}
\label{delta_nu_rho}

Let us first define the large separation ($\Delta \nu$). To this end, it is necessary to introduce the first-order asymptotic relation that permits us to express modal frequencies ($\nu_{n,\ell}$) as a function of the structure of the star, \emph{i.e.} 
\begin{equation} 
\label{asymptotique}
\nu_{n,\ell} \approx  \left( n + \frac{\ell}{2} + \frac{1}{4} + \alpha \right) \left[2 \int_{0}^{R} \frac{\mathrm{d}r}{c_s} \right]^{-1} \, , 
\end{equation}
where $n$ is the radial order, $\ell$ the angular degree, $\alpha$ a term that accounts for the near-surface effect \citep[e.g.][]{JCD97,Rosenthal99}, and $c_s$ the sound speed. Note that such an asymptotic analysis assumes that we consider high radial-order modes \citep[see][]{Tassoul80}. 

Hence, from Eq.~(\ref{asymptotique}), the large separation which is defined as the frequency separation between two consecutive radial orders (for a given $\ell$) is given by 
\begin{equation} 
\label{def_delta_nu}
\Delta \nu \equiv \nu_{n+1,\ell} - \nu_{n,\ell} = \left[2 \int_{0}^{R} \frac{\mathrm{d}r}{c_s} \right]^{-1} \, , 
\end{equation}
and therefore represents the inverse of twice the time for a perturbation of pressure to cross the entire star (in other words a back and forth of a long wavelength pressure wave). 

Now, we aim at deriving the scaling relation between $\Delta \nu$ and the mean density. To this end, one has to adopt the homology relations. So let us consider two stars such that for two shells verifying $r/R = r^\prime/R^\prime$, the corresponding mass shells equal ($m/M = m^\prime/M^\prime$), where $M, M^\prime$ are the total masses of two stars belonging of a homologous series, and $R,R^\prime$ their total radii. This type of approximated stellar models has been extensively discussed in the literature \citep[e.g.][]{CoxGuili68,Kippenhahn90}, and it is possible to show that pressure ($p$) and density ($\rho$) of both models are related by \citep[see Sect. 20.1 of][]{Kippenhahn90}
\begin{eqnarray}
\label{homologous1}
\frac{p}{p^\prime} &=& \left( \frac{M}{M^\prime} \right)^2 \,  \left( \frac{R}{R^\prime} \right)^{-4} \, ,\\
\label{homologous2}
\frac{\rho}{\rho^\prime} &=&  \left( \frac{M}{M^\prime} \right) \,  \left( \frac{R}{R^\prime} \right)^{-3} \, , 
\end{eqnarray}
 Then, from Eqs.~(\ref{homologous1}) and (\ref{homologous2}), the relation of the sound speed immediately follows
\begin{equation}
\label{sound_homologous}
\frac{c_s}{c_s^\prime} =  \left( \frac{M}{M^\prime} \right)^{1/2} \,  \left( \frac{R}{R^\prime} \right)^{-1/2} \, . 
\end{equation}

Now, to demonstrate the relation between the large separation and the mean density, let us define the ratio  
\begin{equation}
\label{ratio}
\mathcal{R} = \frac{\Delta \nu}{\Delta \nu^\prime} \, .
\end{equation}
Using Eq.~(\ref{sound_homologous}) together with the relation $r/R = r^\prime/R^\prime$, it is straightforward to demonstrate the desired scaling relation, \emph{i.e.}
\begin{equation}
\label{ratio}
\mathcal{R} = \left[\int_{0}^{R^\prime} \frac{\mathrm{d}r^\prime}{c_{s}^\prime}\right] \left[ \int_{0}^{R} \frac{\mathrm{d}r}{c_{s}}\right]^{-1} 
=  
\left(\frac{R^\prime}{R}\right)^{3/2}  \left(\frac{M}{M^\prime}\right)^{1/2} = \left(\frac{\bar{\rho}}{\bar{\rho^\prime}}\right)^{1/2} \, .
\end{equation}
In other words, the large separation of a given star ($\Delta \nu$) can be related to its mean density such as 
\begin{equation}
\label{finale_scale1}
\Delta \nu = \left(\frac{\bar{\rho}}{\bar{\rho}_\odot}\right)^{1/2} \, \Delta \nu_\odot , 
\end{equation}
where, as classically found in the literature, the Sun has been used as the reference. Equation (\ref{finale_scale1}) demonstrates the scaling relation between $\Delta \nu$ and $\bar{\rho}$, and shows that the only underlying hypothesis is the homology, which will be discussed in Sect.~\ref{theory_uncertain}. 

At this stage, it is worthwhile to recall that Eq.~(\ref{finale_scale1}) and its derivation is not a novelty and was known, in a hardly different framework, for many decades. Indeed, for classical pulsators, the mode frequencies are often near the fundamental mode frequency (e.g., Cepheid, $\delta$-Scuti, $\beta$-Cephei). For fundamental radial mode ($\nu_0$), its period ($\Pi_0$) is therefore proportional to the time for a pressure perturbation to cross the entire star, \emph{i.e.}
\begin{equation}
\label{fundamental_rho}
\Pi_0 \propto \int_{0}^{R} \frac{\mathrm{d}r}{c_s} \quad \Longrightarrow \quad \Pi_0 \propto \bar{\rho}^{-1/2}
\end{equation}
This relation (Eq.~\ref{fundamental_rho}) is quite famous in the context of classical pulsators and one of the first authors to mention it was, to our knowledge, \cite{Eddington17} for explaining the periodic motion of Cepheids. Subsequently, the relation between $\nu_0$ and the mean density had been often used as an argument to identify periodic motions of stars as pulsations. Moreover, it is the basis of the famous period-luminosity relation of Cepheids \citep[e.g.][]{Cox72,Cox80}. As a conclusion, the derivation of Eq.~(\ref{finale_scale1}) is nothing but the same as for Eq.~(\ref{fundamental_rho}) and was proposed for a long time by several authors such as \cite{Ledoux58}  \citep[see][for a comprehensive discussion]{Cox80}. 

\subsection{Relation between the frequency of the maximum height in the power spectrum ($\numax$) and the cut-off frequency ($\nuc$)}
\label{numax_nuc}

The derivation of the relation between $\numax$ and $\nuc$ is more difficult. Because sometimes not fully understood \citep[e.g.,][]{JCD2011b,JCD2011}, we explain here how the basic physical picture is grasped and that departure from the observed relation arises from the complexity of non-adiabatic processes involving time-dependent treatment of convection and not from the failure of the physical picture. In addition, a discussion on the physics of opacity-driven pulsations in stars will bring us with the conclusion that the main physical reason for the existence of the relation between $\numax$ and $\nuc$ is a common feature of pulsating stars.  

\subsubsection{Derivation of the $\numax-\nuc$ relation}

Let us first begin by recalling that solar-like oscillations are the result from a balance between mode driving and damping. Therefore, as a first approximation, each mode can be considered as a driven and damped oscillator. The driving is related to turbulent Reynolds stresses \cite[see][for a comprehensive review on stochastically excited modes]{Samadi2011}  while damping is caused by a combination of physical processes  not discussed here (but see \cite{Houdek2008} for a discussion on mode damping). 

Therefore, the frequency $\numax$ is in principle determined by both physical mechanisms. To have a more precise view of what governs $\numax$, one has first to determine which of the aforementioned process is responsible for the maximum height in the power spectrum, thus the physical mechanism controlling $\numax$. We thus  consider the height $H$ of a given mode in the power spectrum, which is a natural observable. For stochastically excited modes, the height of the mode profile in the power spectrum is  \citep[e.g.][]{Baudin2005,Chaplin2005,Belkacem2006}
\begin{equation}
\label{heighteqres}
H\;=\;\frac{P\, }{2\,\eta^2\,\mathcal{M}} \, ,
\end{equation}
where $P$ are the excitation rates, $\eta$ the damping rates, and $\mathcal{M}$ the mode masses\footnote{The mode mass corresponds to the total amount of mass effectively moved by a given mode. Its definition is $\mathcal{M} = \int_{0}^{M} \vert \xi \vert^2 {\rm d}m / \vert \xi (r=R) \vert^2$.}. 
It turns out that the depression (plateau) of the damping rates $\eta$ is responsible for the presence of a maximum in the power spectrum. This is in agreement with theoretical computations \citep{Houdek1999,Chaplin2008,Belkacem2011,Belkacem2012} and recent observations of the solar-like stars by \emph{Kepler} \citep{Appourchaux2012}.

The subsequent issue then relies on the physical origin of the depression of the solar damping rates. This plateau originates from a destabilizing effect in the super-adiabatic layers and occurs when the modal period nearly equals the thermal time-scale (or thermal adjustment time-scale) in the superadiabatic layers.  This was first mentioned by \cite{Balmforth92} (see his Sect. 7.2 and 7.3) and confirmed by \cite{Belkacem2011}. The authors used two different non-adiabatic pulsation codes, making this conclusion quite secure.
This can be expressed by the following condition 
\begin{equation}
\label{def_Q}
{\cal Q} =2 \pi \numax \, \tau \sim 1Ê\, , 
\end{equation}
with the inverse of the thermal time-scale defined as  
\begin{equation}
\label{thermal_time}
\tau^{-1} = \frac{L}{4\pi r^2\rho c_v T H_p} = \tau_{\rm conv}^{-1} + \tau_{\rm rad}^{-1} \, , 
\end{equation}
where $L$ is the luminosity, $r$ the radius, $\rho$ the local density, $c_v=\left(\partial U / \partial T\right)_\rho$ with $U$ the specific internal energy, $H_p$ the pressure scale height, $\tau_{\rm rad}$ and $\tau_{\rm conv}$ the radiative and convective thermal time-scales, respectively. Note that Eq.~(\ref{thermal_time}) is a local formulation of the thermal time-scale; a non-local one can also be defined \citep[e.g.][]{Pesnell83}. It is important to note that, contrary to the situation for classical pulsators, for which the envelope is dominated by the radiative transport of energy,  Eq.~(\ref{thermal_time}) exhibits contributions of both the radiative and convective fluxes. 

The last step is to establish the relation between the thermal time-scale ($\tau$) and the cut-off frequency. In the mixing-length theory framework the thermal time-scale can be recast such as \citep{Belkacem2011}
\begin{equation}
\label{def_thermal}
\frac{1}{\tau} = \frac{F_{\rm conv}}{\rho c_v T H_p}  \left[ 1 + \frac{F_{\rm rad}}{F_{\rm conv}} \right] \propto \left(\frac{{\cal M}_a^3}{\alpha}\right) \, \left(\frac{c_s}{2 H_p} \right) \left[ 1 + \frac{F_{\rm rad}}{F_{\rm conv}} \right]  \, ,
\end{equation}
where ${\cal M}_a=\texttt{v}_{\rm conv} / c_s$ is the Mach number, and $\alpha$ the mixing-length, $F_{\rm conv}, F_{\rm rad}$ the convective and radiative fluxes respectively, $c_s/(2H_p)$ the cut-off frequency. Note that the cut-off frequency appears in Eq.~(\ref{def_thermal}) as the ratio $c_s / (2H_p)$.

\begin{figure}[!]
\begin{center}
\includegraphics[height=7cm,width=9cm]{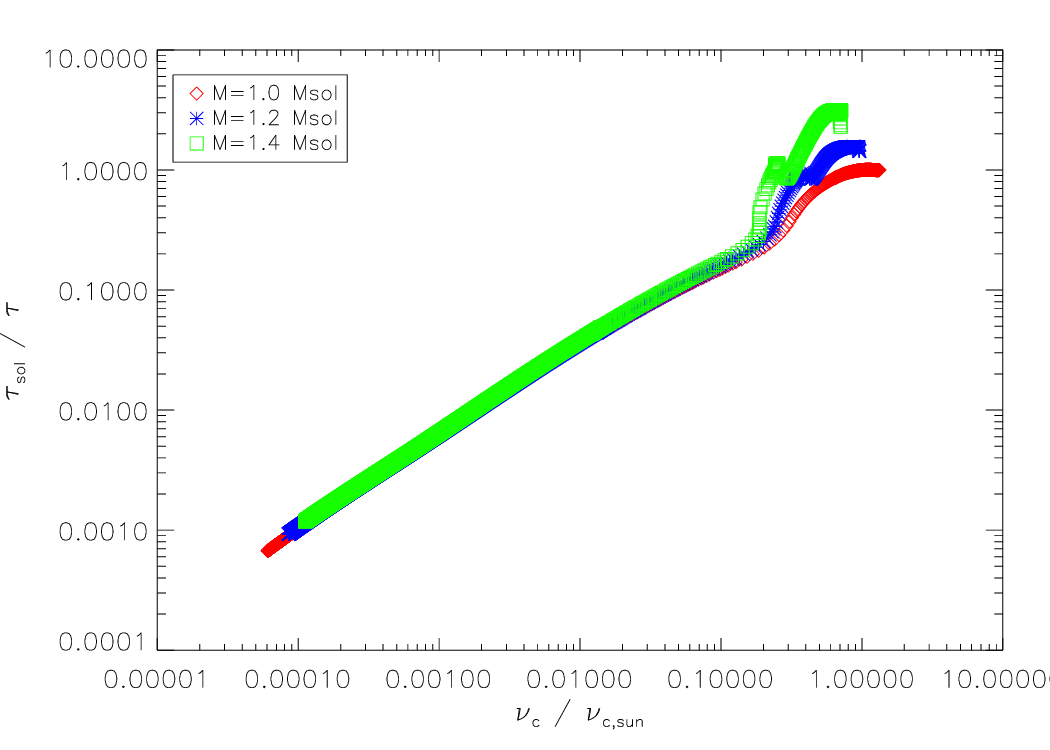}
\caption{Thermal frequency ($1/\tau$) computed from Eq.~(\ref{def_thermal}) versus the cut-off frequency (computed as the ratio $c_s/(2 Hp)$), normalized to the solar values, for models with masses ranging from $M=1.0 \,M_\odot, M=1.2 \,M_\odot$ and $M=1.4 \,M_\odot$ from the ZAMS to the ascending vertical  branch. The inputs physics of the models can be found in Belkacem (2011). }
\label{fig1}
\end{center}
\end{figure}

This relation is verified, in Fig.~\ref{fig1}, by using a grid of stellar models. The relations between the thermal time-scale and the cut-off frequency is very tight for red giants  while a dispersion  is observed for main-sequence stars. Such a dispersion is explained by the dependence to the Mach number to the third power. Indeed, this can be easily understood since the Mach number predominantly depends on the effective temperature that varies much more for main-sequence and sub-giants than for red giants. 
Equation~(\ref{def_thermal}) is useful to explicitly show the relation between $\tau$ and $\nu_{\rm c}$, but   
 an investigation of this relation using  a set of 3D hydrodynamic numerical simulation would be desirable in the future to get more quantitative estimates. 

\begin{figure}[!]
\begin{center}
\includegraphics[height=11cm,width=9cm]{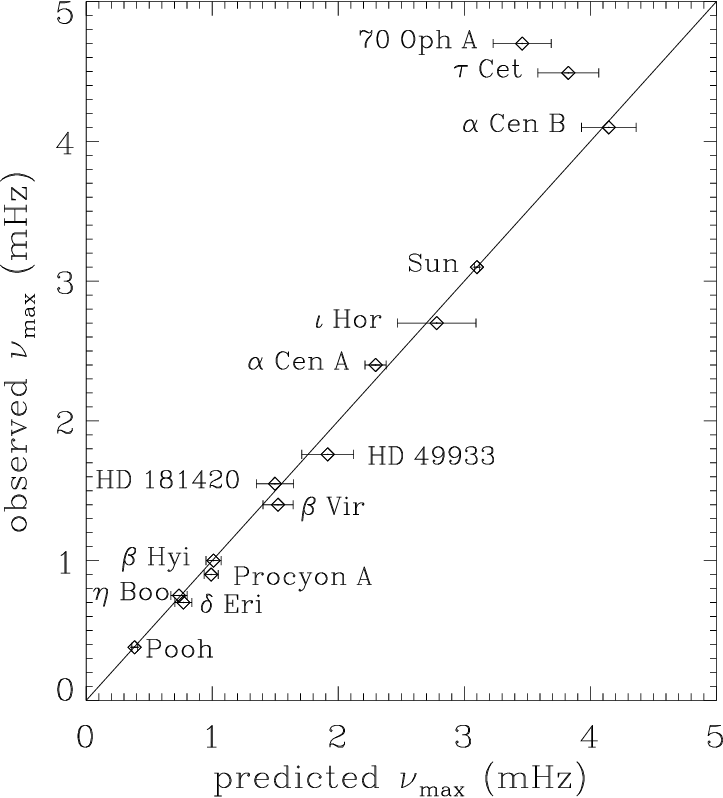}
\caption{Observed $\numax$ versus predicted $\numax$, computed using Eq.~(\ref{final_cut_off}), for stars observed from the ground \citep[see][for details]{Bedding2011}. Figure from \cite{Bedding2011}.}
\label{fig2}
\end{center}
\end{figure}

It is nevertheless worth to emphasize that the resonance between the local thermal time-scale and the pulsation period (which is nothing but the \emph{transition} region, see Sect.~\ref{transition}) explains the \emph{tight}\footnote{Note that the notion of \emph{tightness} sometimes found in the literature is very subjective and would need a more rigorous quantitative estimate.} observed relation between $\numax$ and $\nuc$. More interestingly, as observed in Fig.~\ref{fig2}, the maximum dispersion predicted for main-sequence stars is in agreement with recent observations of \cite{Bedding2011}. Indeed, as depicted by Fig.~\ref{fig2}, a departure from the scaling relation seems to occur for main-sequence stars (i.e., large $\nu_{\rm max}$) in qualitative agreement with Fig.~\ref{fig1}. 

\subsubsection{The transition region: a common feature of both opacity-driven and solar-like pulsators}
\label{transition}

As explained in the preceding section, the cornerstone of the physical picture underlying the $\numax$--$\nuc$ relation is the nearly equality between the thermal time-scale and the modal period in the super-adiabatic layers. 

Let us first briefly introduce the notion of \emph{transition} region \citep[see][for a more complete discussion]{Cox74,Cox80}. Such a region is defined by the layer for which the ratio of the thermal time-scale to the modal period equals unity. Indeed, it separates the quasi-adiabatic from the non-adiabatic layers, more precisely
\begin{itemize}
\item In quasi-adiabatic regions, there is a negligible\footnote{By negligible we mean on the mode stability.} exchange of energy between the oscillations and the background.  In those regions, $\mathcal{Q}\gg 1$ (see Eq.~\ref{def_Q}) then $\tau \gg \Pi_{osc}$,  so that the background does not have time enough to react to the perturbations introduced by one cycle of the oscillation. Consequently, over a time-scale $\tau$ perturbations related to the oscillation compensate and the background experiences a nearly vanishing  average over many oscillation cycles. 
\item In non-adiabatic regions, the energy exchange between the oscillation and the background dominates. We are in the opposite situation where $\mathcal{Q} \ll 1$ so $\tau \ll \Pi_{osc}$. Indeed, in this case the thermal structure "instantaneously" adjusts to the perturbations generated by the oscillation so that, say during compression, any accumulated energy is immediately lost and the perturbation of luminosity is nearly constant.  
\end{itemize}
This transition region is an essential ingredient of the $\kappa$-mechanism in opacity-driven pulsators.  
In those stars, pulsations are destabilized by the perturbation of the opacity. But to be efficient, this destabilization must fulfill several conditions \citep[e.g.,][]{Cox80,CoxGuili68,Pam99} among which the transition region must lie in ionization region. Indeed, the destabilization occurs in the ionization region and if $\tau \ll \Pi_{osc}$ the thermal structure adapts so quickly that the flux is frozen. In the opposite situation if $\tau \gg \Pi_{osc}$ we are in the quasi-adiabatic situation as described above. Finally, one must have $\tau \approx \Pi_{osc}$ in the ionization region for the destabilization to be efficient and dominant over damping terms. 

The situation is similar in solar-like pulsators, except that the destabilization by the perturbation of the opacity never dominates over damping terms and the situation is complicated by the presence of convection which modifies the thermal time-scale (see Eq.~\ref{def_Q}). As shown by \cite{Balmforth92} and \cite{Belkacem2011} the depression of the damping rates occurs when the destabilization by the perturbation of the opacity becomes efficient (and partially compensates the other damping mechanisms), i.e. when the transition region nearly coincides with the ionization region\footnote{Note that in solar-like pulsators, the hydrogen ionization region is located in the uppermost atmosphere and nearly coincides with the super-adiabatic region}. Indeed, for low-frequency modes, $\mathcal{Q} \gg 1$ so that the destabilization is inefficient and, for high-frequency modes, $\mathcal{Q} \gg 1$ the important destabilization in the super-adiabatic layers are compensated by the damping in the atmospheric layers. Therefore destabilization have its maximum impact on the total damping rate for $\mathcal{Q} \approx 1$ in the super-adiabatic layers (near the ionization region of hydrogen), in other words when the transition region occurs in the super-adiabatic layers.   

We then conclude that the observed $\numax$ -- $\nuc$ relation simply results from a common feature of pulsating stars, \emph{i.e.} the occurrence of the non-adiabatic effects in the transition region.

\subsection{Inferring seismic masses and radii from global seismic quantities}

Given the scaling relation described in Sects.~\ref{delta_nu_rho} and \ref{numax_nuc}, it is now possible to link seismic global quantities to stellar parameters. 

Let us start with the scaling relation relating the large separation to the mean density of the star (Eq.~\ref{finale_scale1}), \emph{i.e.}
\begin{equation}
\label{delta_nu_rho_param}
\Delta \nu Ê\propto \bar{\rho}^{1/2} \propto \left(\frac{M}{R^3}\right)^{1/2} \, ,
\end{equation}
where $\Delta \nu$ is the large separation, $\bar{\rho}$ the mean density, $M$ the total mass of the star, and $R$ its total radius. 

The second one relates the frequency of the maximum height in the power spectrum to the cut-off frequency,  \emph{i.e.}
\begin{equation}
\nu_{\mathrm{max}} \propto \nu_{\mathrm{c}}
\end{equation}
The cut-off frequency can be approximatively defined for an isothermal atmosphere \citep[see][for a discussion]{Stello2009} such as
\begin{equation}
\label{cut-off}
\nu_{\mathrm{c}} = \frac{c_\mathrm{s}}{2 H_p} \, , 
\end{equation}
where $c_\mathrm{s}$ is the sound speed. Through the hydrostatic equilibrium, the pressure scale height is related to the gas pressure by $P=\rho g H_p$ and using the ideal gaz equation of state $P\propto \rho T$,  Eq.~(\ref{cut-off}) becomes 
\begin{equation}
\label{final_cut_off}
\nu_{\mathrm{c}} \propto \frac{g}{\sqrt{T}} \propto \frac{M}{R^2 \sqrt{T}} \, .  
\end{equation}
From Sect.~\ref{numax_nuc}, it becomes clear that the introduction of the cut-off frequency is more historical than physically justified. Indeed, \cite{Brown91} conjectured a relation between $\numax$ and $\nuc$ but from a physical point of view it would be more rigorous to mention the relation between $\numax$ and the thermal time-scale  (Eq.~\ref{def_Q}), even if the cut-off frequency can be artificially introduced as in Eq.~(\ref{def_thermal}). 

Now, using Eqs.~(\ref{delta_nu_rho_param}) and (\ref{final_cut_off}), one obtains the desired relations (normalized to the solar values)
\begin{eqnarray}
\label{relations_canoniques1}
\frac{M}{M_\odot} \propto  \left(\frac{\numax}{\numax^\odot}\right)^{3} \, \left(\frac{\Delta \nu}{\Delta \nu_\odot}\right)^{-4} \, \left(\frac{T_{\rm eff}}{T_{\rm eff,\odot}}\right)^{3/2} \, , \\
\label{relations_canoniques2}
\frac{R}{R_\odot} \propto \left(\frac{\numax}{\numax^\odot}\right) \, \left(\frac{\Delta \nu}{\Delta \nu_\odot}\right)^{-2} \, \left(\frac{T_{\rm eff}}{T_{\rm eff,\odot}}\right)^{1/2} \, .
\end{eqnarray}
where $T_{\rm eff}$ is the effective temperature. Note, however, that from Eq.~(\ref{final_cut_off}) to Eqs.~(\ref{relations_canoniques1}) and (\ref{relations_canoniques2}) one implicitly assumes that $T=T_{\rm eff}$. This is not obvious and would deserves further investigations. 

Equations~(\ref{relations_canoniques1}) and (\ref{relations_canoniques2}) constitutes the main frame of what is now commonly called the \emph{ensemble asteroseismology} and recently gives rise to an important work that provide us the tools for a new grip on stellar physics. An all-comprehensive review of the work that make use of the above-mentioned scaling relations would be tedious but it is worth to emphasize the diversity of the applications, namely
\begin{itemize}
\item \emph{Model-independent}\footnote{It would be more accurate to replace the term \emph{model-independent} by \emph{stellar-model-independent} since the derivation of the scaling relations rely on physical assumptions, therefore on a modeling.} \emph{determination of stellar parameters}. Such a type of application is now commonly used to infer masses and radii, so as to identify the observed stars \cite[e.g., ][]{Kallinger2010,Mosser2010}, but  also for other applications such as the characterization of planets \citep[e.g.,][]{Borucki12,Jupiter12}. 
\item \emph{Constraint on stellar evolution and population}. From the determination of stellar parameters for a  large set of stars, ranging from the main-sequence to the red-giant branch, it is  obvious that one of the first  applications is to give constraint on stellar evolution and populations. Such work has been recently performed by many authors among which \cite{Mosser12a} for constraining mass loss at the tip of the red giant branch, or \cite{Miglio09,Miglio12a,Miglio12b} for constraining populations in the milky way. 
\item \emph{A distance indicator}. From the knowledge of stellar radius and with the effective temperature as an input, the distance is derived from a comparison with apparent magnitudes. It is claimed that such a method can provides accurate results, of the order of $10\%$ \citep{Miglio12c,Silva2012}.
\item \emph{Improved determination of $\log g$ and $T_{\rm eff}$}. A striking example of the use of seismic scaling relations is the determination of stellar surface gravities. Classically, surface gravities are obtained by using spectroscopy and isochrone fitting. However, a look at Eq.~(\ref{final_cut_off}) immediately shows that the relation between $\numax$ and $\nuc$ gives access to the surface gravity. It has been shown by \cite{Morel2012} that the seismic gravity is compatible with classical methods and that one can expect a much better accuracy from seismology than from the other methods. Last but not least, the use of this seismic gravity as an input in spectroscopic analysis provides better determination of effective temperature \citep[e.g.][]{Batalha2011,Creevey2012}. 
\end{itemize}

Those few examples offer an overview of the current multiple use of the scaling relations, but ensemble asteroseismology is not limited to these few examples and other scaling relations are promising. 

\section{Some promising scaling relations}

In this Section, we will address three relations that from our point of view are very promising in providing us  highly valuable information for stellar physics, namely the age, luminosity, and effective temperature of stars. 
Those relations are just beginning to be exploited, for different reasons we will discuss below. 

\subsection{The $\Delta \Pi$ -- evolutionary status relation} 

Up to recently, mainly pressure modes have been detected. In contrast, gravity modes (whose restoring force is dominated by buoyancy) were very difficult to detect, especially in the solar case since they have very low amplitudes \citep[e.g.,][]{Belkacem2009,Appourchaux2010}. Nevertheless, the situation changed with the advent of space-borne missions. Indeed, since the first unambiguous detections and  identification of non-radial modes in red giants by CoRoT \citep{DeRidder2009}, a great leap forward is being experienced by the stellar physics community. This has been possible because of the large amplitudes that oscillations develops in red giants compared with the main-sequence stars \citep{Baudin2011a,Samadi2012}. It then makes the detection easier, allowing for the identification and characterization of a very large number of red giants \citep[e.g.,][]{Mosser2010}. 

In main-sequence stars, $p$-modes have high frequencies while $g$-modes have low frequencies. In contrast, in red-giant stars both $p$ and $g$ modes lie in the same frequency range. This is due to the structure of red giants, since the radiative core of red giants contracts from the end of the hydrogen burning phase while the envelope expands. Therefore, the total radius increases, the mean density decreases, then $p$-mode frequencies decrease during the evolution (see Eq.~(\ref{fundamental_rho})). For $g$ modes, we are in the opposite situation since their frequencies increase as the result of the contracting core and more precisely due to the increase of the buoyancy frequency (see Eq.~(\ref{pi_asymp})). 
Then, on the red giant branch, the $p$ and $g$ modes are in the same frequency range, then modes  propagate in both the outer and inner cavities. Moreover, these cavities are coupled by an intermediate zone in which modes are evanescent. Consequently, red giants  exhibit what we call mixed modes \citep[e.g.,][]{Dziembowski2001,Dupret2009}, \emph{i.e.} with a $g$ nature in the core and a $p$ nature in the envelope. As a result, while probing the core they have enough amplitude at the surface of the stars to be detected. Those peculiar physical properties of the oscillations of red giants lead to the detection and characterization of mixed modes in a large sample of stars \citep[e.g.,][]{Bedding2011b,Mosser2011,Mosser2012b}, giving us  the opportunity of a new grip on stellar physics of advanced evolutionary stages. 

\begin{figure}[!]
\begin{center}
\includegraphics[height=9cm,width=13cm]{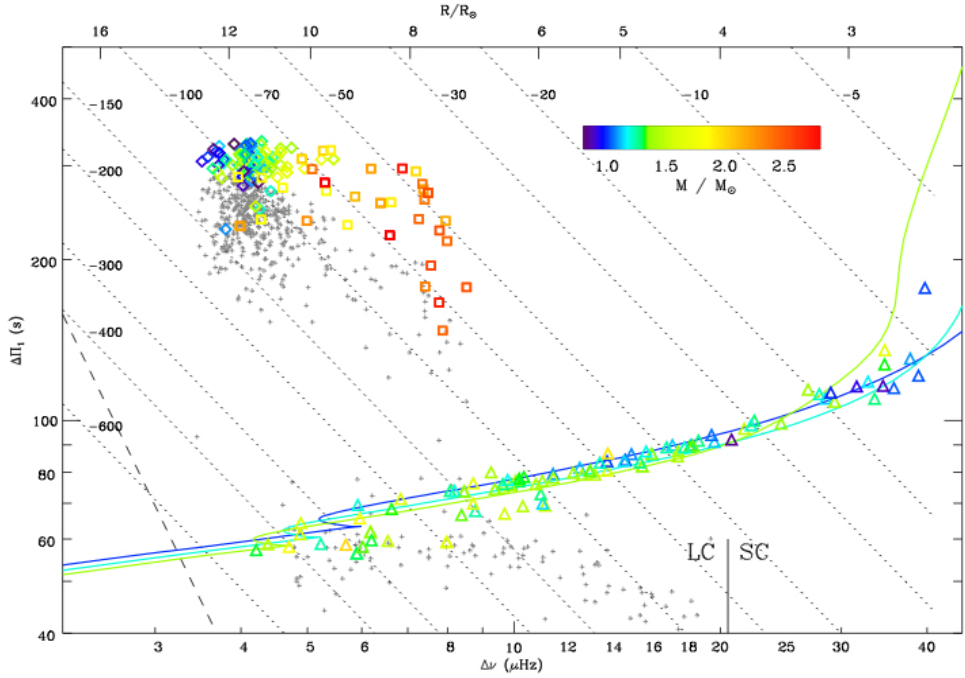}
\caption{Gravity-mode period spacing as a function of the pressure-mode large frequency spacing. RGB stars are indicated by triangles; clump stars by diamonds; secondary clump stars by squares. Small gray crosses indicate the bumped periods measured by Mosser et al. (2011a). The solid colored lines correspond to a grid of stellar models \citep[see][for details]{Mosser2012b}. Figure from \cite{Mosser2012b}. }
\label{fig_benoit}
\end{center}
\end{figure}

To go further, it is first essential to introduce the \emph{period spacing}. It is the counterpart of the large separation for gravity modes. Indeed, in the asymptotic regime, gravity mode periods follow the relation \citep{Tassoul80}
\begin{equation}
\label{pi_asymp}
\Pi_{n,\ell} = \Delta \Pi_\ell (n+\epsilon_g) \, , 
\end{equation}
where $\epsilon_g$ is a phase shift, and $\Delta \Pi_\ell$ the period spacing given  by
\begin{equation}
\label{delta_pi}
\Delta \Pi_\ell = \frac{2\pi^2}{\sqrt{\ell (\ell+1)}} \left( \int_{r_1}^{r_2} N \frac{{\rm d}r}{r} \right)^{-1}
\end{equation}
where $r_1$ and $r_2$ are the radius of the inner and outer turning points of the $g$-mode cavity. It is worth to note that the integral in Eq.~(\ref{delta_pi}) is related to the evolutionary state of the star. Indeed, as the star quits the main-sequence, its core contracts and the buoyancy frequency $N$ increases, leading to a decrease of the period spacing. Therefore, the detection of mixed modes and the period spacing from the observation permitted to assess the evolutionary stage of red-giant stars as shown by  \cite{Bedding2011b} and \cite{Mosser2011}. More precisely, low-mass stars on the ascending red-giant branch (RGB) and after helium ignition (clump stars) can be found at the same location in the HR diagram. Since such stars have similar  envelopes it was impossible to differentiate them from the use of $p$ modes only. In contrast, the detection of mixed modes and the period spacing makes it possible to distinguish between those two evolutionary stages as shown by Fig.~\ref{fig_benoit}. 

Nevertheless, there is a subtlety that cannot be avoided. In fact, the period spacing between two observed mixed modes does not follow exactly the asymptotic relation given by Eq.~(\ref{delta_pi}). This is related to the mixed nature of the mode, since the observed modes are not \emph{pure} $g$-modes they are affected by their acoustic nature and a departure from Eq.~(\ref{delta_pi}) follows. To cope with this issue, \cite{Goupil2013,Mosser2012b}, proposed an asymptotic relation adapted for mixed modes (based on the formalism developed by \cite{Shibahashi79}), \emph{i.e.}
\begin{equation}
\label{asymp_mixed}
\nu_{m} = \nu_{n_p,\ell=1} + \frac{\Delta \nu}{\pi} \arctan \left[ q \tan \pi \left( \frac{1}{\Delta \Pi \; \nu} - \epsilon_g \right) \right] \, , 
\end{equation}
where $\nu_{m}$ is the mixed mode frequency, $\nu_{n_p,\ell=1}$ the frequency of \emph{pure} $p$ modes, and $q$ a coupling factor. Therefore, using Eq.~(\ref{asymp_mixed}) with the measured $\nu_{m}$ yields a determination of the period spacing ($\Delta \Pi_\ell$). This is a crucial step for comparing the observations and the modeling as well as for determining the evolutionary stage, as illustrated by Fig.~\ref{fig_benoit}. 
 
Finally, one can conclude that the scaling relation between the period spacing and the evolutionary state of stars is only in its infancy but still very promising since it provides unprecedented information of the innermost layers of stars. 

\subsection{Scaling relations related to mode amplitudes and linewidths}

Asteroseismology is not limited to the analysis of mode frequencies but also to mode amplitudes and linewidths. The latter being related to the exchange of energy between the oscillation and the background,  and not only to the star structure, they are subject to the uncertainties related to the coupling between convection and pulsation. Nevertheless, several successful attempts have been proposed to scale those seismic parameters to the stellar fundamental parameters. 

\subsubsection{Mode amplitude \emph{vs} $L/M$}

We first consider the relation between mode amplitudes and stellar parameters, both in terms of velocity and intensity fluctuations. 
On the basis of the theoretical calculations of \cite{JCD83b}, \cite{Kjeldsen95}  derived the first example of a scaling relation given in terms of the maximum of the mode surface velocity (hereafter $V_{\rm max}$). This scaling predicts that 
\begin{equation}
V_{\rm max} \propto \left(\frac{L}{M}\right) \, ,
\end{equation}
where $L$ is the luminosity. 
\cite{JCD83b} assumed that there is an \emph{equipartition} between the energy carried by the most energetic eddies and the modes. As mentioned by \cite{Belkacem2009} and \cite{Samadi2011}, a necessary (but not sufficient) condition for having such an equipartition is that turbulent viscosity is the dominant source of damping. However, there is currently no consensus as to what is the dominant physical processes contributing to the damping of $p$-modes.  

Prior to the CoRoT mission, observations of mode velocity in solar-like stars were sparse \citep[e.g.][]{Houdek2002} but motivated several theoretical work on the physical mechanisms underlying mode driving \citep{Houdek1999,Houdek2002,Samadi2007,Samadi2011}. With the launch of space-borne mission CoRoT, and its unprecedented high-quality data, such work on scaling relation of mode amplitudes became achievable \citep{Michel2008,Baudin2011a}. This has been confirmed with the {\it Kepler} mission, which motivated an important work on those relations \citep[e.g.][]{Kjeldsen2011,Huber2011,Mosser12a}. 

The large amount of high-quality data from CoRoT and {\it Kepler} led to a variety of results that paradoxically complicated the picture since no clear scaling relation emerged. 
\cite{Samadi2012} addressed this issue for red giants, based on theoretical developments, a set of 3D numerical simulations, and the observations. 
They found that the maximum amplitude (in term of velocity) follows the scaling relation
\begin{equation}
V_{\rm max} \propto \eta_{\rm max}^{-1/2} \left(\frac{L}{M}\right)^{1.3} \left(\frac{M}{R^3}\right) ^{0.525} \, ,
\end{equation}
where $\eta_{\rm max}$ is the linewidth at $\nu=\nu_{\rm max}$. In terms of intensity fluctuations, \cite{Samadi2012} showed the necessity to go beyond the adiabatic relation between velocity and intensity, especially for red giants, and proposed
\begin{equation}
\left(\frac{\delta L}{L}\right)_{\rm max} \propto \eta_{\rm max}^{-1/2} \left(\frac{L}{M}\right)^{1.55} \left(\frac{M}{R^3}\right) ^{0.5} \, .
\end{equation}
These scaling relations show a systematic discrepancy for red-giants, which is attributed to non-adiabatic effects. Consequently, we can state that the scaling relation of mode amplitudes is now well understood for main-sequence stars but still need to be investigated for red giants. 

\subsubsection{Mode linewidths \emph{vs} $T_{\rm eff}$}

For mode linewidths (or equivalently mode damping rates), scaling relations have been investigated only very recently. This is the result of the need for long-time and almost-uninterrupted monitoring to resolve individual modes and to enable their precise measurements. 

\begin{figure}[tbp]
\begin{center}
\includegraphics[width=11cm]{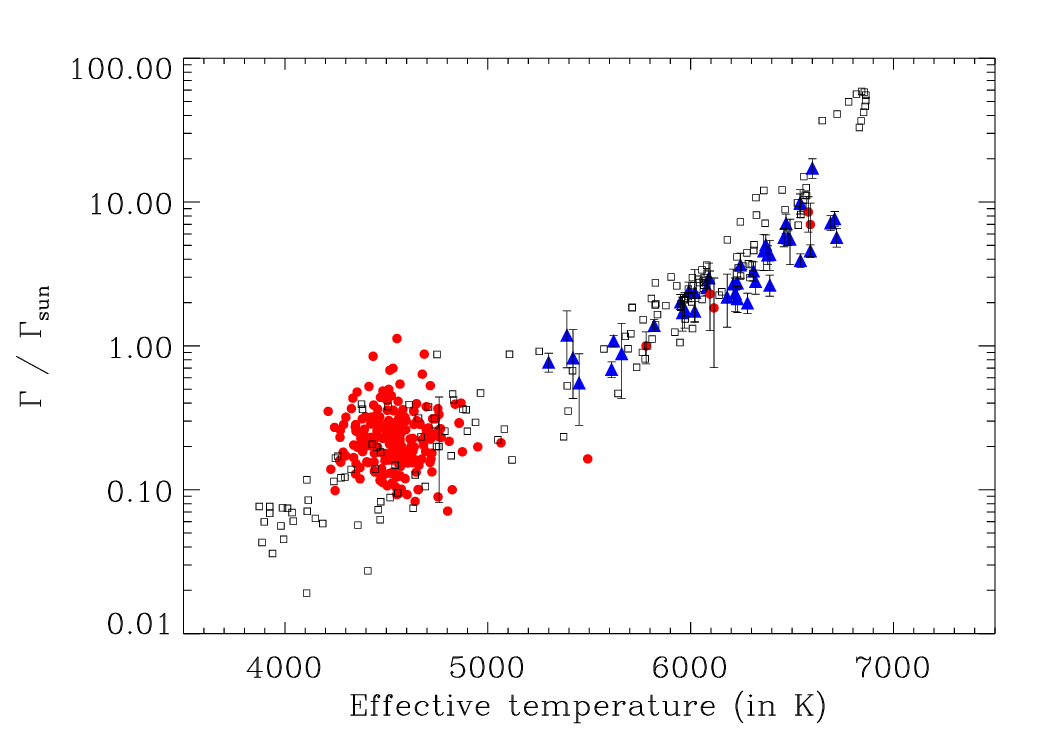}
\caption{
Mode linewidths (normalised by the solar value, $\Gamma_{\rm sun} = 0.95 \, \mu$Hz) versus effective temperature. The squared symbols represent theoretical calculations computed as explained in \cite{Belkacem2012}. The triangles correspond to the observations of main-sequence stars  derived by \cite{Appourchaux2012} from the {\it Kepler} data (with their 3-$\sigma$ error-bars). The dots correspond to the observations of red giants (with $T_{\rm eff} < 5200$ K) and main-sequence  (with $T_{\rm eff} > 5200$ K, with their 3-$\sigma$ error-bars) stars as derived by \cite{Baudin2011a,Baudin2011b} from the CoRoT data. }
\label{Kepler_Corot}
\end{center}
\end{figure}

\cite{Houdek1999}, and later \cite{Chaplin2009}, have investigated the dependence of mode-damping rates on global stellar parameters. From  ground-based measurements, \cite{Chaplin2009} found that observed mode linewidths follow a power-law  of the form $\eta \propto T_{\rm eff}^4$ (where $T_{\rm eff}$ is the effective temperature) and no clear tendency emerged when $\eta$ is scaled with the ratio $L/M$. 
Nevertheless, these measurements were based on short-term observations and derived from an  inhomogeneous set of analysis and instruments, resulting in a large dispersion. This was settled by \cite{Baudin2011a,Baudin2011b}  (Fig.~\ref{Kepler_Corot}) using a homogeneous sample of CoRoT data. They found that a unique power-law hardly describes the entire range of effective temperature covered by main-sequence and red-giant stars and proposed that mode linewidths of main-sequence stars follow a power-law of $T_{\rm eff}^{16 \pm 2}$, while red-giant stars only slightly depend on effective temperature ($T_{\rm eff}^{-0.3 \pm 0.9}$). The latter result was later confirmed and extended by {\it Kepler} observations (Fig.~\ref{Kepler_Corot}) to main-sequence and sub-giant stars \citep{Appourchaux2012}. We also note that  
\cite{Corsaro2012} proposed that mode linewidths follow an exponential power law. In absence of a strong theoretical argument to adopt either a power law or an exponential, the statistical significance must dictate our choice and this is still to be performed. From a theoretical point of view, \cite{Chaplin2009}, based on the formalism developed by \cite{Balmforth92,Houdek1999} and \cite{Chaplin2005}, predicted a power-law of $\eta \propto T_{\rm eff}^4$ which disagrees with CoRoT and {\it Kepler} observations \citep{Houdek2012}. 
In contrast, \cite{Belkacem2012}, based on the formalism of \cite{MAD05}, were able to reproduce both CoRoT and {\it Kepler} observations. 

To get more insight into the relation between $\eta$ and $T_{\rm eff}$, let us first write down the integral expression of the damping rates  \citep[e.g.][]{MAD05}
\begin{equation}
\label{damping}
\eta = \frac{1}{2 \, \omega I} \int_{0}^{M} \mathcal{I}m 
\left[ \frac{\delta \rho}{\rho_0}^* \frac{\delta P}{\rho_0} \right] \textrm{d}m \, , 
\end{equation}
where $\omega$ is the mode frequency, $\delta \rho$ the Lagrangian perturbation of density, $\delta P$ the perturbation of the total pressure (including the turbulent pressure), $\rho_0$ the mean density, and where the star denotes the complex conjugate. The mode inertia is given by 
\begin{equation}
\label{inertia}
I = \int_{0}^{M} \left\vert  \vec \xi \right\vert^2 \textrm{d}m \, , 
\end{equation}
where $\vec \xi$ is the eigendisplacement vector. 
Therefore, a look at Eqs.~(\ref{damping}) and (\ref{inertia}) makes it clear the need for disentangling the effects of mode inertia and the work integral (\emph{i.e.}, the integral appearing in Eq.~\ref{damping}). 
The latter is related to non-adiabatic processes corresponding to a transfer of energy between pulsation and convection. Hence, it can be assumed at first glance that the work integral scales dimensionally with the ratio $L/M$\footnote{This comes from the perturbed energy equation that shows that entropy perturbation dimensionally scales as the ratio $L/M$.}. As verified by \cite{Belkacem2012}, it follows that the relation
\begin{equation}
\label{etaI}
\eta \, I \propto \left( \frac{L}{M} \right)^{2.7} \, .
\end{equation}
holds. In contrast, the mode inertia ($I$) does not depend on mode energy leakage but on the star's static structure\footnote{Except through non-adiabatic effects on mode eigendisplacement.}, and more precisely on the properties of its uppermost layers. Hence, one can expect mode inertia to scale with the surface gravity\footnote{Note that mode inertia also scales with the dynamical timescale $\sqrt(GM/R^3)$ with almost the same dispersion as for the surface gravity.}. It has been shown in \cite{Belkacem2012} that 
\begin{equation}
  \label{Inertie}
  I \propto g^{-2.4} \, .
\end{equation}
Using Eq.~(\ref{etaI}) and Eq.~(\ref{Inertie}), it turns out that 
\begin{equation}
\label{eta_modif}
  \eta \propto T_{\rm eff}^{10.8} \; g^{-0.3} \, .
\end{equation}
This simple analysis allows us to explain qualitatively the strong dependence of mode damping rates on effective temperature. Finally, it is important to stress that Eq.~(\ref{eta_modif}) is compatible with current {\it Kepler} observations (T. Appourchaux, private communication). 

\section{Do seismic scaling relations give us accurate stellar global parameters?}

This is a difficult and still open question issue. We explain here what are the main sources of uncertainties. The latter includes biases, all unknown and missing processes, as well as the precision of the measurements. 

The typical precision of the stellar seismic indices is very good (better than $1 \%$ for $\Delta \nu$ and $5 \%$ for $\numax$ in most cases). Under the assumption that the effective temperature is determined with a precision of, say,  $100$ K, it translates into a precision of  $20 \%$ for the mass, $8 \%$ for the radius, and $0.04$ dex for $\log g$. It is worth to mention that the precision on $\numax$ is the limiting factor. These numbers are very encouraging and often presented as the uncertainty, but they do not include biases. 
Indeed, efforts are currently undertaken to assess the reliability of those numbers and more precisely to identify the biases from both the observational and the theoretical side.

\subsection{Observational uncertainties}
\label{biais_obs}

To illustrate the importance of considering the possible sources of biases with great care let us consider two examples, namely $\Delta \nu$ and $\numax$. 

\subsubsection{Observational determination of $\Delta \nu$}
\label{delta_nu_obs}

As demonstrated by \cite{Verner2011} on the basis of hare-and-hound exercises and fitting of {\it Kepler} data by six teams, the way the large separation is determined plays an important role. Indeed, the authors conclude that the expected relative precision on $\Delta \nu$ is about $2\%$. 

We also emphasize that the large separation is a quantity that is relevant in an asymptotic regime only. Thus, it can be extracted if the assumption underlying the asymptotic analysis is respected, here the requirement is that one must consider high radial-orders. However, in practice, solar-like oscillations are observed for low to moderate values of the radial orders, especially for red giants ($n \in [6;15]$), which make the basic assumption violated. This is a non-negligible source of biases as demonstrated by \cite{Mosser2013}, and can generate systematics as high as of $8 \%$ for the mass and $4 \%$ for the radius determination. 

\subsubsection{Observational determination of $\numax$}
\label{numax_obs}

The second example concerns the determination of $\numax$, for which several issues arise. Since we are dealing with a stochastic process, the maximum of height in the power spectrum can vary depending on the considered time-duration of the observations. This makes the determination of $\numax$ quite unsecured and requires very long observations to settle the problem (unfortunately this is not always possible) and to   approach the stationary state. Therefore, except if we are able to demonstrate that stochastic effects are negligible, it is safe to consider that a possible bias of about half a large separation. The related uncertainties can be roughly estimated by considering the ratio $\Delta \nu / (2 \numax)$. It is about 2.5\% for main-sequence stars and 5\% for red-giants. Note that such an effect has been considered for red giants by \cite{Mosser2011b}. 
In addition, as for $\Delta \nu$, several methods exists for deriving $\nu_{\rm max}$ from the light-curve and \cite{Verner2011} found an average precision of about $4 \%$. 

Moreover, another issue naturally arises;  
{\it can $\numax$ be observationally determined by both considering the maximum of amplitude or height in the power spectrum?} Following the work of \cite{Belkacem2011} the suitable choice (in the sense it is physically grounded) is to use the maximum height. Indeed, the relation between $\numax$ and $\nuc$ is due to the occurrence of the transition region in the superadiabtic layers that translates, from an observational point of view, by the depression of the damping rates. Hence, the quantity one must consider must be dominated by the damping processes. It is the case for mode height, but not for mode amplitude. Indeed, mode amplitude derives from a more subtle balance between mode driving, damping, and mode inertia. Therefore, the maximum of height and amplitude is expected to be different. 

\subsection{Theoretical uncertainties and additional dependencies}
\label{biais_theo}

As illustrated in Sect.~\ref{biais_obs}, biases can arise from the observational determination of seismic indices. However, as shown in Sects.~\ref{delta_nu_rho} and \ref{numax_nuc}, scaling relations are based on modeling and therefore cannot be considered as perfect, so does the derived stellar parameters. 

In the following, we discuss briefly the main physical reason underlying the dispersion of those relations. 

\subsubsection{The limits of the homology assumption and the $\Delta \nu$ -- $\bar{\rho}$ relation}

\begin{figure}[tbp]
\begin{center}
\includegraphics[width=11cm]{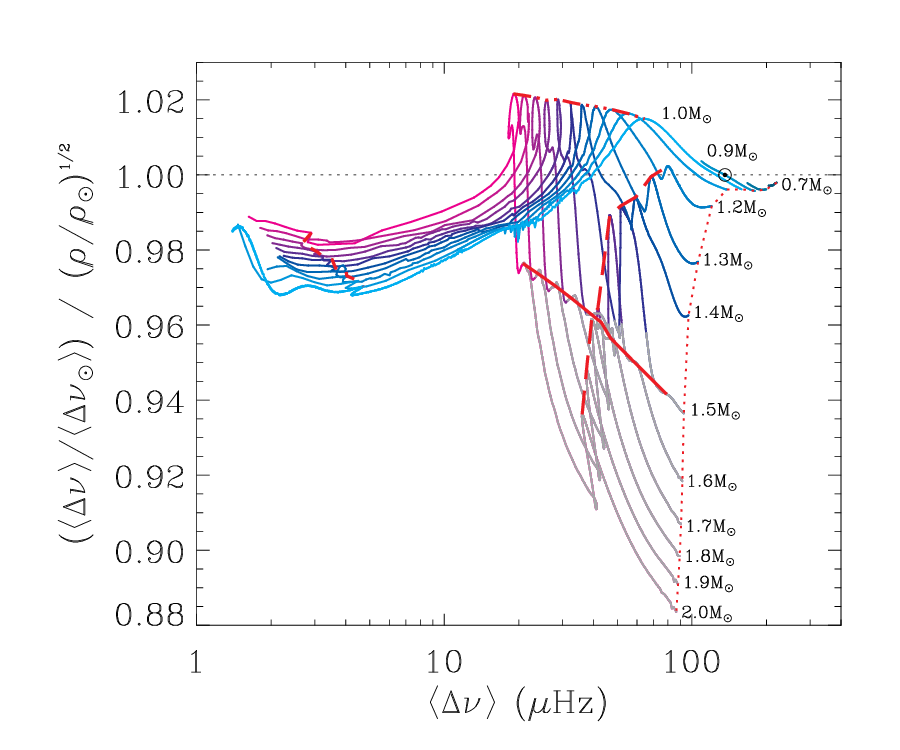}
\caption{Ratio between the large separation and the squared mean density of stars, normalized to the solar values, from the ZAMS to the ascending red-giant branch \citep[see][for details]{White2011}.  Figure from \cite{White2011}.}
\label{white}
\end{center}
\end{figure}

As depicted in Sect.~\ref{delta_nu_rho}, the physical assumption that permits us to relate the large separation to the mean density of a star is homology. While often considered as a crude approximation to derive the internal structure of a star, this approximation nevertheless gives ones quite a good result concerning this scaling. Figure~\ref{white} illustrates that; for typical solar-like oscillating stars, this relations holds within 3 to 4\%. 

To understand such a departure, it is necessary to recall the main physical hypothesis that make the homology strictly valid. The first main requirement is hydrostatic equilibrium, \emph{i.e.} that the acceleration term in the momentum equation must vanish. The second one is thermal equilibrium, \emph{i.e.} that the energy generated must be strictly compensated by the energy emitted (in other words, ${\rm d}S/{\rm d}t = 0$, where $S$ is the entropy). Finally, as demonstrated by \cite{CoxGuili68}, the homology requires that the constitutive equations (such as opacity, production rate of energy, etc...) must be power-laws of their arguments. 

It is clear that during its evolution, a star breaks all the requirements implying a departure from the homology. 
This scaling relation would then need a deep investigation of the physical reason explaining precisely the origin of the departure from the homology. This requires to consider each evolutionary state (main-sequence, sub-giant, and red-giant phase) separately since the physics differs from one to another. Such a work is highly desirable in the future, to understand why this relation is so precise and to propose improvements. 

\label{theory_uncertain}

\subsubsection{Influence of the Mach number to the $\numax$ -- $\nu_{\rm c}$ relation}

An other example is the relation between $\numax$ and $\nu_{\rm c}$, which is not exact as shown in Sect.~\ref{numax_nuc}. More precisely, the relation between the thermal frequency and the cut-off frequency is not direct and many other physical quantities appear, among which the Mach number is the more important. As already explained in Sect.~\ref{numax_nuc}, the cut-off frequency is artificially introduced for historical reasons and this subsequently leads us to introduce the Mach number. 
We recall that this number is the ratio of the turbulent (or convective) velocity to the sound speed and permits us to measure the \emph{degree} of turbulence of a flow. 

In solar-like stars, this number is almost negligible in the interior ($\mathcal{M}_a \ll 1$) but increases significantly near the photosphere. In these layers, convection becomes inefficient and convective velocities increase rapidly over a relatively small radial scale to sustain the convective flux. As a result, in this region the Mach number reaches a maximum (which is of the order of 0.3 for the Sun). From one star to another, this number varies typically from 0.3 to 0.7, mainly depending on the evolutionary status and the mass of the considered stars \citep{Houdek1999}.  
 
Therefore, from Eq.~(\ref{def_thermal}), it becomes clear that such an extra-term in the relation between $\numax$ and $\nuc$ is to be investigated. Simple theory of convection such as mixing-length theories, aside from giving very different results for the convective velocities \citep{Samadi2006}, provides us unrealistic estimates of turbulent velocities \citep[e.g.][]{Samadi2003}. Therefore, the only way to overcome this difficult problem is to use 3D hydrodynamical simulations to get more insight into the evolution of the Mach number across the HR diagram (work in progress). 

\section{Concluding remarks}

In this review, we have discussed the now commonly used scaling relations ($\numax$--$\nuc$ and $\Delta \nu$--$\bar{\rho}$) for deriving seismic masses and radii. We have shown that their physical justifications were already known, for a long time, in the context of classical pulsators. Therefore, it is striking to note that those relations derive from common features of pulsating stars and not solar-like pulsators only. 
We also discussed how the seismic masses and radii are derived and emphasized the numerous and important applications for improving our knowledge of stellar structure and evolution. We did not limited our discussion to these scaling relations but also addressed other scaling relations we consider as promising, namely; the relations between mode amplitudes and luminosity, between mode linewidths and effective temperature, and between period spacing and evolutionary status of stars. 

We then focus our discussion on the uncertainties related to those scaling relations. It turns out that uncertainties arise from both the observational and theoretical sides. Consequently, the stellar parameters as derived from the scaling relations also suffers from those uncertainties. Therefore, one of the crucial step to obtain precise and accurate seismic stellar parameters is the calibration. Preliminary work in the direction has been achieved \citep[e.g.][]{Silva2012,Huber2012}, and give encouraging results. 

\begin{acknowledgements}
I am grateful to M.J. Goupil, B. Mosser, and R. Samadi for reading the manuscript and for many fruitful  discussions. I also thank T. Bedding and T. White for providing Figs.~\ref{fig2} and \ref{white}.
\end{acknowledgements}


%
\end{document}